# Indian Cosmological Ideas

Roopa Hulikal Narayan

## 1. Introduction

This paper, third in the series on Indian tradition of physics, describes conceptions of the cosmos. The first two papers [2], [6] described the Nyaya-Vaisheshika school and its theory of matter, together with the consideration of absence of matter. A part of this tradition was to consider the nature of the cosmos, and although not so clearly described in the foundational sutra texts, these ideas are clearly spelt out in texts such as Yoga Vasishtha (YV). We will sketch these ideas and draw parallels with the similar conceptions that have arisen in modern physics.

The specific questions that will be examined will be the place of the observer and the nature of the universe. We will begin with the concept of *abhava* in Kanada's sutras and examine its implications. We will validate our interpretations by means of textual references from YV. In particular, the conception of multiple universes that occurs often in YV will be examined in the framework of the Indian physics. The other surprising concepts that are discussed in YV include flow of time and its variability with respect to different observers, and the possibility of passage across universes (worlds). Although these ideas are best seen as speculative scenarios, they reveal a lot about the nature of the physical ideas of that time. We argue that the basis of these ideas goes back in the Kanada's sutras.

As background, there appear to have been several models of the universe current in ancient India, which were of varying abstraction. The model of Yajnavalkya [5], [6] seems to be at the basis of medieval speculations on the nature of the universe. Although these models do not address the question of "absence of matter" explicitly, it is implicit elsewhere.

## 2. Philosophical background

Kanada on whose philosophy many of the ideas in this paper are based was a realist. He as quoted above believed that there is no authority which need be accepted without question. Vasishtha, like Kanada, also saw the limitations of logic and how it leads to paradoxical situations. Although YV is often considered an expression of an idealist approach to reality, there are other more realistic interpretations of it.

The YV text has both the observer and the material viewpoint in it. The philosophical arguments are based in the materialistic interpretation of the cosmos. The universe which is created also suffers decay and complete dissolution.

According to Patanjali, in a given universe there is a constant tendency of things to go back to the original state of equilibrium which constitutes of assimilation and dissipation - just the opposite of the process of differentiation in the integrated i.e., the initial integrated mass energy was differentiated to the specific/*vishesha* to evolve in to the universe. The mass energy total of the universe is always being dissipated however slowly or imperceptibly which is dissolving in the original formless *prakriti* which is a permanent state of equilibrium from which there is no return. This is the statement of increase in entropy in the universe [3].

Therefore in general it is the same energy force which goes through different cycles of creation and may be manifested as any form of matter. This is the reason Indian philosophers invoke the concept of *maya*. This must not be confused as a denial of the reality of the existence of a universe or the life forms on it, but understood as just a way of stressing that there is no permanent reality and change defines the order of things.

## 3.1 *abhava* as the non- *bhava*

Vacuum is generally understood as the empty space which is interspersed in the universe we live in along with matter which may be in different forms. A more expansive definition of vacuum may be arrived at through Kanada's description of *abhava*. He states:

*abhava* is भावभिन्न = Other than *bhava*/existence.

Kanada's universe as described in his sutras mainly consists of the describable [6] which is the *bhava*. This 'other' cannot refer to non-perceptible objects because mind and soul are categorized as those beyond perception and yet *abhava* is not part of that.

तत्रात्मा मनश्चाप्रत्यक्षे ॥८।१।२॥

There in, the mind and soul are non-perceptible 8.1.2 [7].

Kanada's world of matter has enumerated non-perceptible entities like mind, soul and infinite entities like space, time and if *abhava* is that which does not belong to any of these categories, it indeed can only be defined as that which is other than all the known. The method of understanding matter as described by Kanada is knowing the unknown based on comparison with what is known through their respective similarities and differences. Since the sutra format used by Kanada is very concise and it never states more than necessary, it is strange that Kanada should describe *abhava (*vacuum) as that which is other than *bhava* instead of ignoring it, as the only universe which is real should be what we perceive.

In this context his definition of 'existent/*bhava*' is helpful:

सदिति यतो द्रव्यगुणकर्म्मसु सा सत्ता ॥१।२।७॥

Existence is that to which are due the belief and usage, namely '(it is) existent', in respect of matter, attribute and action. 1.2.8 [7].

In the next nine sutras that follow he states that existence/ *bhava* is neither the matter nor its properties or attributes or any of the other classifications. Therefore he must be stating it as a way of understanding the world we live in.

The power or energy of infinite consciousness is always said to be in motion overlooking all epochs of time which endows all matter with characteristic quality and this power is called *niyati* or the great existence. This is compared with movement of water which gives rise to whirlpools [4].

There are three different kinds of *abhava* explained [6] and the later thinkers have interpreted *abhava* to mean vacuum or nothingness alone that exists in the pre-production state and post-destruction state of matter.

## 3.2 Creation or Dissolution

A different interpretation of Kanada's *abhava* supported with other texts is arrived at here.

His first sutra on non-existence/*abhava* is as follows:

क्रियागुणव्य☐देशाभावात् प्रागसत् ॥६।१।१॥
Due to prior non-applicability of action and attribute it is the opposite of existence (*abhava* )  9.1.1 [7]

In this sutra the state of existence at a state when it has no attributes is referred to as *abhava*. Kanada describes matter as the inherent cause of attribute and action [6] at an early stage of his text which has led to arguments about what happens to matter when it is attribute-less. At that stage in his text he continues with discussions about matter in all details.

This concept of *abhava* is at the end of the text and there are separate sutras addressing material objects, soul, etc. He waits till he has established the definition of *bhava*/existence as that which is understood as the visible matter and the instruments of perception in all possibility in this world. Even in this sutra he emphasizes the attribute and action which implies that by stating *abhava* as the opposite of such pre-established *bhava* he means that which is something other than what is understood in this specific fashion of attributable properties.

Kanada states in the first part of his book that there are only six predicable and no more. So is he only describing this world with its contents which are perceivable through the observer here and since this is what he is describing as *bhava* will the opposite of it refer to what is from other worlds or universes? This is supported by the sutra 1.2.8 as well.

He also theorizes that the *anu*/atoms don't lose their attributable nature during chemical reactions or any other process except at the time of cosmic dissolution, though the properties of an end-product of a reaction varies from those of the reactants since the resultant product which is a new form of matter must be differentiable through properties from the initial reactants [2].

Kanada comes from a tradition where the cosmological ideas were well established and most of the ancient works and rituals of India have cosmological significance which are almost always stated in a very coded way as seen from today's context. Therefore there is a further need to examine this possibility.

It is mentioned even in Yoga Vasistha that the five basic elements which constitute matter cannot be resolved back in to these basic elements except during the cosmic dissolution [4, Page 48-50].

Indian cosmology has always made its objective to understand creation at three levels as *vyakta*/the manifested, *avyakta*/un-manifested and *purusha*/observer and Kanada begins his work by stating that he intends to describe the *vyakta*/the manifested [1]. This further goes to say that it is this world alone he intends to describe through his predicable.

Therefore the kind of *abhava* referred here has to be that which precedes the existing universe in time which is the cosmic creation or a later point of time in the universe which is the cosmic dissolution. The countless universes are continuously created and destroyed [4, Page 26] without the knowledge of each other's existence [4, Page 66], which means this process of attributable matter reaching non-attributable state is part of a recursive cycle of *bhava to abhava.*

## 3.3 The Cycle

It is not a surprise that Kanada's next sutra is

सदसत् ॥६।१।२॥
Existent becomes non-existent 9.1.2

Kanada after describing *abhava* as the non-attributable state of matter, in this sutra he explains that such a state is indeed attained by the matter. He is not describing a process of material object being created and destroyed as interpreted by many scholars in this world as is observed by us but he is also talking of what may not be observed.

The Indian physicists are apparently aware of conservation of matter and state that the order of universe remains unaltered by anything, for something can never become nothing [4, Page 66]. The conservation of matter or energy (explained later) is the basic law of nature in Indian cosmology.

Even in the Upanishads it is stated that the world in the beginning was of the category of non-existence. The Samkhya system of India and the Puranas too talk of a cyclic universe [1].

Therefore Existent becoming non-existent refers not to material object's physical destruction, but instead it is the Cosmic dissolution of the universe which is the *abhava* mentioned here. This sutra with the previous one describes one complete cosmic cycle of matter universe from the cosmic creation to the cosmic dissolution.

The cyclic system of the universe is given a definite life time period of 8.64 billion years although there are cycles with longer as well as shorter life terms [11].

## 3.3 The Samkhya

असतः क्रियागुणव्य☐देशाभावादर्थान्तरम् ॥६।१।३॥
Non-Existence has a difference in meaning from existence due to the non-existence of a reference of it through action and attribute. 9.1.3 [7]

Kanada couldn't have got any clearer than this in explaining *abhava* as something which is other than what we see as the perceived world we live in through certain attributes and properties. The properties and attributes are identification coordinates of matter in this world, but now it is obvious that the universe itself is more than that.

If non-existence meant nothing, then Kanada should have said that a non-existence which cannot be described in his terms has no meaning instead of the fact that it differs in meaning.

The Samkhya, Yoga, Nyaya and Vaisheshika are Schools held together in terms of their basic philosophy of the Cosmos in accord with the Vedas although their specific definition and understanding of matter in academic details may vary. Samkhya too expresses the basic *gunas*/attributes as those which help distinguish the posterior and the prior from the middle which is the matter activity. Here a clear stand is taken on the posterior and prior referring to the cosmic creation and dissolution process leaving no room for doubt. The matter universe is also a gradient to distinguish the *avyakta*/un-manifested and *purusha*/observer [1].

This is further supported by the next sutra:

सच्चासत् ॥६।१।४॥
Existent also is Non-existent. 9.1.4 [7]

Kanada in this clarifies the fact that it is the matter universe which is also the kind of *abhava* which precedes the formation of matter and is consequent of its destruction.

## 3.4 Cosmic Energy

The Puranas, Samkhya, etc describe the universe as being created and absorbed back in to a kind of ground-stuff reality. The only category which transcends all paradoxes and oppositions and is ground stuff reality is called the cosmic energy [1].

The YV describes space, time, light and all other elements of universe being created from the cosmic energy and the real universe that we perceive is inherent in it [4, Page 46-50]. This cosmic energy is everything and the opposite of everything. In it everything is negated [5]. It covers all the definitions of *bhava* and *abhava* as explained in the sutras of Kanada which go through cycles of inter-conversion.

Hence Kanada must be referring to the cosmic creation, dissolution and the energy as understood in cosmology since in this sutra he explains absolute non-existence, in a different place he describes *abhava* as vacuum and therefore the only other kind of *abhava* which is an offshoot from Kanada's own definition of *bhava* is that of other universes from the one we live in.

## 4.1 Multiversity

This cosmology assumes that an infinite numbers of universes co-exist each in a different state of formation or destruction without the knowledge of each other [4, Page 60]..

The Vedas and texts nearer to the Vedas chronologically also mention many universes co-existing [1]. In Yajnavakya's Brihadarnyaka Upanishad which is an early text and supposed to be the origin of India astronomy the entire cosmic space is filled with many egg shaped universes each outside the other. The inter-planetary distances of our own solar system are calculated. The infinite number of universes is an accepted concept even in the *puranas* which are contemporary with Aryabhata [8], [13], [14].

The period of life cycle of universe is given and the speed of light is calculated as 4,404 *yojanas* per *nimisha* which is very closely 186,000 miles/sec [11]. The diameter of the universe is given to be 500 million *yojanas* with other universes around it [11].

Millions of universes are said to appear in the cosmic space like specks of dust in a beam of light with their components like space, time, action, matter, day and night [4, Page 120]. But just like our planet with indigenous variety of life, all these universes too have similar beings with different bodies sited to those universes [4, page 146]. These multiple universes are not even governed by the same laws as ours or did not even evolve similar to our universe [5]. This may also be the reason why the human mind is considered incapable of conceiving the infinite existence outside our own world.

## 4.2 Comprehension of the Multiverse

Can the numerous universes which cannot be described through the properties of matter in our world be visualized or comprehended in any way at all? In Yoga Vasisitha it is said that like all the attributes are inherent in matter the experiences of innumerable universes is inherent in an observer [4, Page 261] and inseparable from her.

The comprehension is done through the mind of the observer and the nature of observer in Indian tradition is worth understanding.

## 4.2.1 Nature of the Observer

In the tradition of Samkhya and Vaisheshika, the understanding of an observer is explained at two levels:

i. mind as an instrument of understanding
ii. an inhered understanding or awareness of ground-stuff reality which supersedes the matter notion of universe. This notion of the observer demands that the observer be present for the universe to exist as is which is in agreement with the modern physics [1].

The mind of an observer is said to have no existence independent of the cosmic mind. The cosmic mind simply means the cosmic energy or the basic energy which is the fundamental constituent of all the universes in all their states.

These other universes are said to be experienced in a unique way which is subjective like the dream objects are experienced only by the dreamer. The objects of perception as experienced by the perceiver are also unique [4, Page 40].

The role of the observer is very central to the understanding of all these universes in Indian tradition which is done through the categories of mind although it is different from the way an observer is understood in the modern physics. The Samkhya is very acute and developed in explaining the external physical universe as reflected in the inner structure of mind and body [10]. A logical analysis of the working of the mind is done thoroughly which we shall briefly examine. Time and again it is insisted that only because the observer's mind is made of the same energy as the infinite power, the observer inheres the ability to experience the cosmos [4].

## 4.2.2 Mind and Perception

Perception through the sense-organs is the direct cognition [9].
This sense-perception occurs in two stages:

1. *Nirvikalpaka*/Abstract Perception
The relevant sense-organ of the observer comes in contact with the object of perception and there arises a non-determinate perception; by which is meant that perception which is free from all notions of name, genus and such other details which objectifies the mere thing in itself as 'something' in a vague form.

2. *Savikalpaka*/Determinate Perception
The objectification of the object to be perceived is followed by a definite perception of the object with respect to the attributes and the attributed, qualities and qualified. This results in the naming of the object as a specific *padartha*/predicable. A deeper study of the object leads to classification of the same in to respective genus, species, etc.

An object is perceived only after an objectification of the perceived object abstractly takes place in the mind. Such an objectification of multiple universes is probably what is said to be inhered in the mind or to which a predisposition exists.

The relation between the observer and the observed is very important in cosmology. In YV it is said thought is inherent in the mind and the object of perception is inherent in the perceiver. The author says -Who has ever discovered a distinction between the two? [4, Page 41]. The two are indistinguishable because without the perceiving mind the perceived remains meaningless. The complementarity's between the outer world and the inner world of the observer is of fundamental significance [10]. To understand one in the absence of other is meaningless since one helps define the other.

The *nirvikalpa samdhi*/abstract perception state, where physically one remains in the same place but mentally reaches a different plane is used to explain time and space traveling. From a physical point-of-view there exists a vast separation between the planes [4, Page 56-57]. The cosmic energy, mind and infinite space are all said to be made of the same substance pervaded by the all pervasive cosmic energy [4, Page 91]. Here the abstractness may be a reference to the fact that the cosmos with multitude of universes can be understood only when one dissociates oneself from the understanding of existence in this world.

In conclusion, the knowledge of cosmic energy which transcends the categories of space, time and matter can be obtained by fine tuning the mind to it since all realities are topography of the mindscape though such knowledge will be in terms of the previous associations of the mind [11].

## 5. 1 Time as Cause

Hours and minutes are an external measurement of time based on the revolutions of sun but it is not inhered in time itself i.e. it is not an innate property of the infinite cosmological time. Here Time is only a mathematical determinant quality in the course of measuring the existence of an existential thing [7].

Kanada defines time as an independent definable [6]. It is characterized as eternal and a unity [6]. It is *asamavayi*/non-inhered in any object. It exists as the cause of the universe.

Time as a causal agent is interpreted by Prashastapada as its all pervasive nature. Time has a similar reference in other texts. It alone creates innumerable universes and in a very short time destroys everything. Its essential nature is hidden. Time is established in the absolute being or the cosmic energy. Even though Time creates endless universes, it remains a constant. Time destroys everything but yet remains undestroyed.

The other aspect of time as that which is responsible for birth and death is always acknowledged.

In this context Kanada's following sutra must be re-examined.

तत्त्वम्भावेन ॥२।२।८॥

Unity (of Time is explained) by (the unity of) existence.  2.2.8 [7]

In explaining the unity of time the reason for Time appearing as manifold is given as *upadhi*/external condition imposed by the movement of sun, etc. A separate explanation is given for the three-fold past, present and future of time as a limitation of the determinant.

Perhaps Kanada say manifold instead of three-fold for he is discussing the varying time with respect to other universes. Time is unaffected even by the fires of cosmic dissolution. Time therefore is causal for all the universes present through their numerous dissolutions and formations. It is said to be an entity which cannot be analyzed [11]. This may be the context in which even the infinite cosmos is said to be incomprehensible.

### 5.2.1 Time flow

The flow of time is indicated as varying with the universes as indicated in many stories. In this section mainly stories from the YV are considered though similar instances are found in the Puranas, Mahabharatha, Upanishads.

**The Infinite Cosmos**

The infinite objective creation is said to resolve in to one homogenous form called Brahman or consciousness during cosmic dissolution yet retaining the potentiality and basic qualities of creation. The observer or mind is not independent of this. The perceived object is inherent in this observer. After a cosmic dissolution and before the next epoch begins the entire objective universe is in a state of perfect equilibrium.

Countless universes emerge from this state of equilibrium similar to infinite sunrays from the sun. Even this state is said to be infinite and since infinite emerges from infinite to exist as infinite, the world is said to be never created. This is to state that energy is conserved in effect and hence there is no true creation as such. There are fourteen planes of existence each with its own types of inhabitants. All the different species in the infinite universes are with bodies suitable for that world [4].

One cannot understand consciousness as long as one is used to looking at the world of matter and when this outlook is overcome it is possible to see the entire cosmos as essentially the same kind of energy [4, Page 56].
The different universes are said to be in different stages of evolution and to travel from one to the other one needs to go through a subtle hole. Vasistha insists that time, space, duration and objectivity do not emerge from matter for that would make them material-like.

The observer is the individualized cosmic consciousness and all the universes coexist without being aware of mutual existence similar to all the people on earth are not aware of every human's existence.

**Time flow and travel**

Three kinds of space are categorized as the infinite space which is probably the outer cosmos, the psychological space which is the mind and hence the observer, and the physical space which must be the world we perceive. Travel in the infinite space by the observer leads to a different world which is in a different frame of time. The time flow in some of these worlds is said to be much slower than our world which is explained as a result of time and space not having a fixed span.

The epoch in a universe which is just the twinkling of an eye in another universe is said to be like real-time when in that universe similar to a huge object reflected in the small sized mirror.

An individual from one world traveling to another with a different time-span lives there in the new time-frame and on returning to her original world finds herself at a time which has moved on by several human life cycles. This is amazingly similar to the twins' paradox problem where the space traveler finds herself younger than the twin back home.

This is the mathematical concept of infinity as applied to time and the infinite energy from which infinity can emerge or dissolve in it and yet the ground stuff remains unaffected. Also an awareness of time as applied today at both microscopic and astronomical ranges exists. Time-travel which is also space travel is mentioned as both a psychological and physical process.

A greater understanding of the Time travel demands an in-depth study of it by itself. Here it is sufficient to recognize that as a concept time-travel existed in Indian Cosmology. There are many more stories with many instances of traveling from one universe to another through the aid of infinite space, mind and Cosmic Energy.

**Space**

Space is infinite and eternal with a non-atomic structure. There are three types of space – *akasha*/the psychological space, *paramakasham/*the physical space and *brahmakasham*/the infinite space. The infinite space is the most subtle which exists in the middle when the finite intelligence travels from one place to another, for it is infinite [4].

In the three-fold space, the infinite space is the absolute, the finite space of the observer creates the division of time and pervades all beings, and in the physical space the material world exists [4].

Therefore space and time do not have a fixed span [4, Page 55] that means that they cannot be absolute. They do not arise from matter, for they then would be material [4] instead they are said to emerge from the cosmic energy.

In the Mahabharata the infinite Space is said to be non-measurable even with self-luminous universes [11]. Every thing in all universes are necessarily in the infinite space which is not escapable even for the thought [4].

The infinite Space is spherical in nature. Billard, Seidenberg, Subhash Kak and others have argued that not only did ancient Indians have a tradition of scientific astronomical observation, but also the works like Surya Siddhanta and the Vedas themselves are based on a mathematical pattern of the universe. The spherical/*parimandalya* nature of space, time and atoms is therefore a fundamental concept of the Indian cosmos rather than a cosmetic reference to a physical shape [1]. There are several instances where there is travel between different universes with different time-frames.

## Conclusion

In Indian tradition universe is cyclic and going through continuous cycles of formation and destruction with intermediate times of brief rest which is recursive. Everything in the cosmos is interrelated by a common substratum which is called the *Brahman*/energy which is conserved. Quantum non-locality too leads some physicists to theorize a fundamental holistic connection among all particles of the universe [15].

Time and space are infinite with no fixed span and time and life forms vary with the universe. The importance of the awareness or consciousness of an observer is a primary difference with the current science and this observer cannot understand the outer cosmos through the perceived matter in this world.

The universe is subjective to the individual mind which is a reflection of the Cosmic mind which may be comparable to the fact that everyone cannot understand the paradoxes of quantum mechanics or relativity. A diagrammatic representation is as follows:

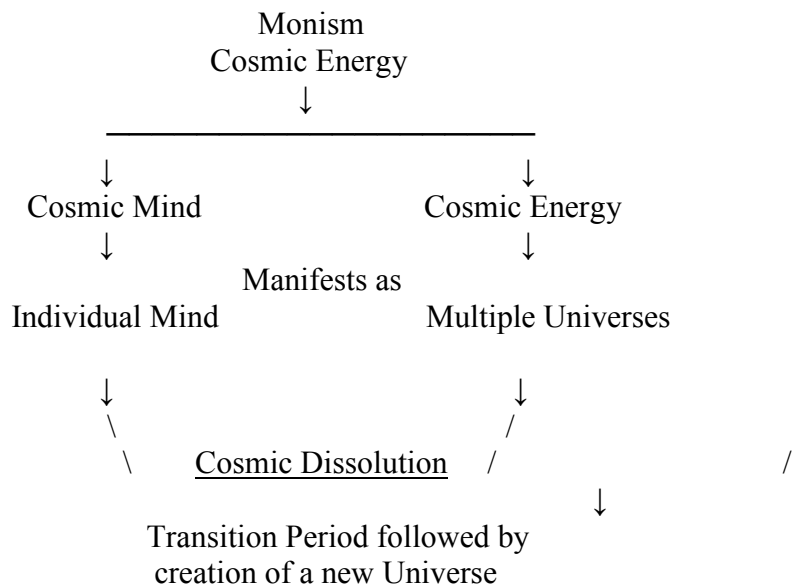

A dualistic experience is observed when the cosmic energy differentiates in to the individual mind which is the observer and the matter universe which is the observed.